\documentclass[aps,superscriptaddress,twocolumn,showpacs]{revtex4-1}
\pdfoutput=1
\usepackage{color}
\usepackage{amsmath, amsthm, amsfonts}    
\usepackage{amssymb}
\usepackage{mathtools}
\usepackage{mathptmx} 
\usepackage[table]{xcolor}
\usepackage{dsfont}
\usepackage{enumitem} 
\usepackage{appendix}
\usepackage{braket}
\usepackage[skins,theorems]{tcolorbox}
\tcbset{highlight math style={enhanced,
  colframe=red,colback=white,arc=0pt,boxrule=1pt}}
\usepackage{cancel}
\usepackage{empheq}
\usepackage{lipsum}
\usepackage{graphicx}
\usepackage{tikz}
\usepackage{makecell}
\usepackage[unicode=true,pdfusetitle,
 bookmarks=true,bookmarksnumbered=false,bookmarksopen=false,
 breaklinks=true,pdfborder={0 0 0},backref=false,colorlinks=true,citecolor=blue]{hyperref}
\usepackage{graphicx}   
\usepackage[caption=false]{subfig}       
\usepackage{bm}            
\usepackage[normalem]{ulem}  

\newlength\imagewidth
\newlength\imagescale
\def\be{\begin{eqnarray}}
\def\ee{\end{eqnarray}}
\def\r{{\bf r}}

\def\E{{\bf E}}

\def\im{{\rm i}}

\DeclareUnicodeCharacter{2212}{-}

\definecolor{JOT-color}{named}{blue}
\definecolor{CSF-color}{named}{orange}

\newcommand{\rr}{\color{red}}\rr
\newcommand{\bb}{\color{black}}

\begin{document}
 
\title{Spheres of maximum electromagnetic chirality}




\author{Jorge Olmos-Trigo}
\email{jolmostrigo@gmail.com}
\affiliation{Centro de F\'isica de Materiales, Paseo Manuel de Lardizabal 5, 20018 Donostia-San Sebastian, Spain.}

\author{Manuel Nieto-Vesperinas}
\affiliation{Instituto de Ciencia de Materiales de Madrid, Consejo Superior de Investigaciones Científicas. Campus de Cantoblanco, Madrid 28049, Spain.}

\author{Gabriel Molina-Terriza}
\affiliation{Centro de F\'isica de Materiales, Paseo Manuel de Lardizabal 5, 20018 Donostia-San Sebastian, Spain.}
\affiliation{Donostia International Physics Center, Paseo Manuel de Lardizabal 4, 20018 Donostia-San Sebastian, Spain.}
\affiliation{IKERBASQUE, Basque Foundation for Science, Mar\'ia D\'iaz de Haro 3, 48013 Bilbao, Spain.}

\begin{abstract}
The search for objects that yield maximum electromagnetic chirality in their emitted wavefield has garnered significant attention in recent years. However, achieving  such maximum chirality  is challenging, as it typically requires complex chiral metamaterials. Here we demonstrate that chiral spheres can yield maximum  chirality in their emitted wavefield. Specifically, we analytically find the spectral trajectories at which chiral spheres become optically transparent to a given helicity of the incident field, while for its opposite helicity, they behave as dual objects, i.e., on scattering, they preserve helicity.  Since chiral spheres 
behave as dual objects at the first Kerker condition of zero backscattering, we significantly simplify  this condition in terms of a Riccati-Bessel function. Importantly, all our results are exact and applicable regardless of the multipolar order, refractive index contrast, optical size, and intrinsic chirality of the chiral sphere. Thus, our exact findings can serve as building blocks for designing novel metasurfaces or metamaterials with maximum electromagnetic chirality properties. 
\end{abstract}

\maketitle

Chiral objects cannot be superimposed with their mirror images by translations or rotations~\cite{petitjean2003chirality, hentschel2017chiral, chen2022can}. 
Many crucial organic molecules, such as Glucose, DNA, and most biological amino-acids, are chiral.  
Nowadays, it is  recognized that chirality plays a vital role in the chemical interactions of chiral drugs in the human body regarding their curative potency and toxicity~\cite{chen1982quantitative}.  For instance, the thalidomide tragedy demonstrated that any drug that manifests in enantiomers (mirror pairs of chiral molecules) must be considered as two different drugs whose efficacy and side effects must be treated and determined separately~\cite{lenz1962thalidomide}. Consequently,  detecting  and characterizing enantiomers has been of utmost relevance in the biomedical and pharmaceutical industries~\cite{mcbride1961thalidomide, nugent1993beyond, kim2022enantioselective}.  
Besides its geometry, another feasible way to infer
the chirality of an object is based on its interactions with  chiral light. 

In electromagnetism, the most usual  magnitude to unravel the chirality of a  sample  is the handedness of circularly-polarized (CPL) fields~\cite{tang2010optical, sanz2020artificial}, commonly referred to as helicity~\cite{calkin1965invariance}. 
In addition to the potential applications of helicity in the context of optical forces~\cite{cameron2014discriminatory, canaguier2013mechanical, ding2014realization, tkachenko2014helicity, nieto2015optical,
hayat2015lateral, liu2018polarization,
li2019optical,
nieto2021reactive, genet2022chiral, sifat2022force, wu2022selective}, spin-orbit interactions of light~\cite{marrucci2006optical, bliokh2015spin, olmos2019enhanced, olmos2019asymmetry, forbes2019spin, guo2019routing, mullner2022discrimination, gautier2022planar}, or its fundamental connection to Kerker conditions~\cite{ zambrana2013duality,  olmos2020unveiling, PhysRevApplied.18.044007}, 
helicity can also be used to probe chiral and achiral objects~\cite{yoo2015chiral, nieto2017chiral, poulikakos2019optical, crimin2019optical, olmos2023capturing}.  Now, an object  is electromagnetically achiral if all  components of the scattered field under the illumination with CPL electromagnetic waves of one sign can also be mimicked with incident CPL waves of the opposite sign.  Otherwise, the object  is electromagnetically chiral~\cite{hanifeh2020helicity}. 

However, chiroptical responses are  typically weak; thus,  the intrinsic chirality of samples is often not detected~\cite{tang2010optical}, and researchers  have recently turned their attention to the search of objects of maximum electromagnetic chirality (MECh) ~\cite{fernandez2016objects}. In short, all chiral samples that are transparent to all fields of one helicity are objects of MECh~\cite{fernandez2016objects}. Additionally, if such object is reciprocal,  the previous argument is bidirectional: all objects of MECh are optically transparent to a given helicity of the field. Moreover,  reciprocal objects of MECh preserve the opposite helicity of the incident field upon interaction. That is, such objects have electromagnetic duality as a result of the fact that they are reciprocal~\cite{fernandez2016objects}.

However, to date,  objects yielding MECh have only been unveiled in complex chiral metamaterials. 
For instance, in a double-turn helix~\cite{doi:10.1021/acsphotonics.1c01887}, on metasurfaces made of pairs of dielectric bars~\cite{gorkunov2020metasurfaces}, in dielectric metamaterials consisting of a random colloid of meta-atoms~\cite{asadchy2020three}, or in a photonic crystal  with a chiral array of perforating holes~\cite{semnani2020spin}, among others~\cite{hoflich2019resonant,gorkunov2021bound,kuhner2022unlocking,lim2022maximally, zhao2022realization,shi2022planar}.  
In this regard, we should note that numerical methods are  involved in all previous examples devoted to the search for objects yielding MECh ~\cite{doi:10.1021/acsphotonics.1c01887,gorkunov2020metasurfaces, asadchy2020three, semnani2020spin,hoflich2019resonant, gorkunov2021bound,kuhner2022unlocking,lim2022maximally, zhao2022realization,shi2022planar}.  This is because the constitutive relations of chiral media have only been analytically solved  for the case of spherical objects by Craig Bohren in 1974~\cite{bohren1974light}.  However, whether spherical chiral particles  may behave as  objects of  MECh  remains  unexplored~\cite{shang2013analysis, hu2022achieving}.

In this work, we demonstrate that  chiral spheres can behave as objects of MECh. Based on energy conservation, we first derive the equation that such  a  sphere satisfies. Shortly after, we solve this equation by analytically finding the spectral trajectories at which chiral spheres are  optically transparent to a given helicity of the incident field. In contrast, for the opposite helicity of the incident field,  they behave as dual objects, i.e., on scattering, they preserve  helicity as a consequence of reciprocity for MECh objects~\footnote{Note that chiral spheres are by construction reciprocal, and thus, meet the Onsager relations. As a consequence of reciprocity,  if a chiral sphere is transparent to a given helicity of the incident field, then the opposite helicity of the incident field is necessarily preserved upon interaction.}. 
Moreover, since electromagnetic duality is a characteristic of spheres also  when they meet the first Kerker condition (K1) of zero backscattering \cite{nieto2015opticaltheorem}, 
we disclose the analytical requirements of K1  for chiral spheres  in terms of  a Riccati-Bessel function. This offers a straightforward, systematic procedure to establish the K1 condition. 
Note that  the so-called generalized Kerker effects~\cite{liu2018generalized,alaee2015generalized, bag2018transverse, shamkhi2019transverse} do not generally correspond to dual particles and, thus, do not emerge for spheres of MECh. 

We first address magnetodielectric  chiral spheres of high refractive index (HRI), which, as far as we know, have not yet been synthesized; subsequently,
we present results for so far existing dielectric chiral spheres of lower refractive index.

Our findings  are exact and remain valid regardless of the  multipolar order, refractive index, optical size, and intrinsic chirality. Spheres yielding MECh  can be used as building blocks of novel  maximum electromagnetic chirality nanoantennas, metasurfaces,  and metamaterials. Therefore, our predictions suggest experimental observations.


In what follows, we will derive the general conditions that must be imposed on the scattering coefficients for a sphere to become an object of MECh. They can be summarized with a simple formula:
\be
\label{General_spheres_maxchiral}
\boxed{
\; a_\ell  = b_\ell  = i \sigma c_\ell  \; \forall \ell \quad \text{and} \quad c_\ell \neq 0 .
}
\ee
The implication of Eq.~\eqref{General_spheres_maxchiral} is that for the incident helicity $\sigma = \pm 1$, we get $\tilde{a}_{\ell,\sigma}=\tilde{b}_{\ell,\sigma}=0$, (cf. Appendix~\ref{th}), while for the helicity $-\sigma$, $\tilde{a}_{\ell,\sigma}=\tilde{b}_{\ell,\sigma}= 2 i c_\ell$~\footnote{Note that the theoretical framework is introduced in Appendix.~\ref{th} while the explicit form of the $a_\ell$, $b_\ell$, annd $c_\ell$ chiral Mie coefficients  are given in~\ref{Bohren}. }. Note that $\{\tilde{a}_{\ell,\sigma} , \tilde{b}_{\ell,\sigma}  \}$ are given by Eq.~\eqref{effective}. 
As a result, the scattering cross-section for any incident field will be zero for helicity $\sigma$, while it will be non-zero for an incident field carrying the opposite helicity. 

To start the derivation, let us use the GLMT in order to compute the extinction and scattering  cross-sections, $Q^\sigma_{\rm{ext}}$ and $Q^\sigma_{\rm{sca}}$, respectively. We  generalize the work by Gouesbet et al., devoted to  the GLMT for achiral spheres~\cite{gouesbet1985scattering}, to obtain 
\be \label{ext}
k^2 Q^\sigma_{\rm{ext}} &=& \sum_{\ell =1}^{\ell = \infty} |C^{\sigma}_{\ell  m}|^2 \Re \{\tilde{a}_{\ell,\sigma} + \tilde{b}_{\ell,\sigma}   \}, \\  \label{sca}
k^2 Q^\sigma_{\rm{sca}} &=& \sum_{\ell =1}^{\ell = \infty} |C^{\sigma}_{\ell  m}|^2 \left( |\tilde{a}_{\ell,\sigma} |^2+ |\tilde{b}_{\ell,\sigma} |^2 \right).
\ee
Moreover, let us also address the absorption cross-section,   $Q^\sigma_{\rm{abs}} = Q^\sigma_{\rm{ext}} - Q^\sigma_{\rm{sca}}$, which we will use later on.

The derivation of Eq.~\ref{General_spheres_maxchiral} proceeds in the following way: 
\begin{enumerate}
    \item [1)] We will show that objects of MECh  must be lossless.
    \item [2)] Then, we shall show that an antidual sphere must be completely transparent, i.e., the extinction cross-section is identical to zero.
    \item [3)] We will apply 1) and 2) to a chiral sphere and see that the conditions for MECh  are those of Eq.~\eqref{General_spheres_maxchiral}.
\end{enumerate}

\begin{figure*}[t!]
  \includegraphics[width=\textwidth]{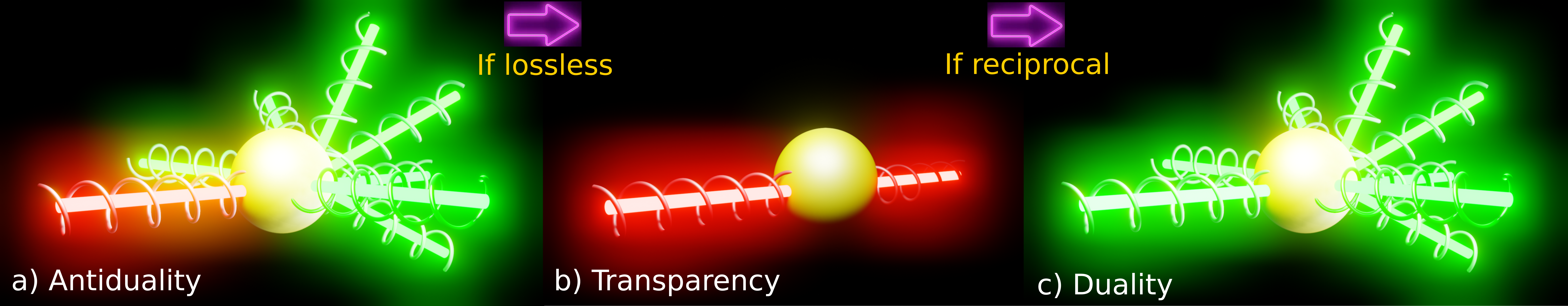}
  \caption{\color{black}{Scattering features of a   sphere of maximum electromagnetic chirality  under the illumination of a well-defined helicity field. Red and green colors correspond to left and right circularly polarized fields, respectively.  a) Antiduality condition (helicity is sign-flipped after scattering). b) Optical transparency (no scattering). c) Duality condition,  i.e., preservation of helicity after scattering, which is also a characteristic of the first Kerker condition, which leads to zero backscattering for cylindrically symmetry objects.  The left panel highlights that optical transparency is reached for an antidual chiral sphere if and only if it is also lossless. The right panel highlights that duality is reached for reciprocal spheres of MECh upon flipping the helicity content of the incident field.}}
  \label{F_1}
\end{figure*}

To start with, we observe that objects of MECh  are optically transparent to a given helicity of the incident field~\cite{fernandez2016objects}. Therefore, such objects cannot absorb  photons with that helicity. In this regard, Kirchhoff’s thermal radiation law states that the emissivity of a body  must be equal to its absorptivity~\cite{kirchhoff1860g, planck1914theory,greffet}. In other words,  a lossy generic object  must emit. Therefore, to avoid emission, the object must necessarily be without losses. 


Now, we turn to the derivation that an antidual sphere must be completely transparent. By definition, such sphere must not scatter light with the same helicity as the incident field~\cite{zambrana2013dual}. Let us impose optical transparency in the same helicity channel as the incident field ($\sigma = \sigma'$), namely, $| \E^{\sigma \sigma }_{\rm{sca}}( \r) | = 0$ (see Appendix~\ref{th}). Then from Eqs.~\eqref{sca_1}-\eqref{sca_2}, we straightforwardly note that 
\begin{align} \label{trans_main}
\tilde{a}_{\ell,\sigma}  = -\tilde{b}_{\ell,\sigma}  \; \forall \ell   &\Longleftrightarrow a_\ell  + b_\ell  = 2i \sigma c_\ell  \; \forall \ell    &\text{for}& \; \; \sigma' = \sigma. 
\end{align}
From Eq.~\eqref{ext}, we  note that Eq.~\eqref{trans_main} is a {sufficient condition} for total transparency to a given incident helicity  since then the extinction cross section goes to zero. That is, imposing partial transparency in the same helicity channel as the incident field ($\sigma = \sigma'$)   naturally leads to $Q^\sigma_{\rm{ext}}=0$.
Note that in this derivation, we have employed the formulas for a chiral sphere to apply them to the next part of the derivation directly. Although complete transparency of an antidual sphere is a general result~\cite{olmos2020optimal, olmos2020kerker, lasa2023resonant}, when applied to chiral spheres, if they are total transparent for a given incident helicity, they may be non-transparent for the opposite incident helicity.

It is crucial to notice that the non-absorbing version of  the conservation energy ($Q^\sigma _{\rm{ext}} = Q^\sigma _{\rm{sca}}$), together with Eq.~\eqref{trans_main}, also imposes $Q^\sigma _{\rm{sca}} = 0$. Now, according to Eq.~\eqref{sca}, the only possible solution is then given by $\tilde{a}_{\ell,\sigma}  = \tilde{b}_{\ell,\sigma}  = 0$, which leads to~\footnote{Note that this argument is bidirectional. That is, if $\tilde{a}_{\ell,\sigma}  = \tilde{b}_{\ell,\sigma}  = 0$, then $Q^\sigma _{\rm{sca}} = 0$.}
\be
\label{chiral anapoles_main}
\boxed{
\tilde{a}_{\ell,\sigma}  = \tilde{b}_{\ell,\sigma}  = 0 \; \forall \ell  \Longleftrightarrow a_\ell  = b_\ell  = i \sigma c_\ell  \; \forall \ell .
}
\ee

This ends our derivation, as these equations lead directly to the requirements we anticipated in Eq. (\ref{General_spheres_maxchiral}). Therefore, while for incident helicity $\sigma$, $\tilde{a}_{\ell,\sigma}  = \tilde{b}_{\ell,\sigma}  = 0$, for the opposite helicity of the incident field, we obtain $\tilde{a}_{\ell,\sigma}=\tilde{b}_{\ell,\sigma}= 2 i c_\ell$. As expected and due to the fact that chiral spheres are reciprocal objects~\cite{onsager1936electric}, $\tilde{a}_{\ell,\sigma}=\tilde{b}_{\ell,\sigma}= 2 i c_\ell$ preserves the helicity content of the incident field.

Equations~\eqref{General_spheres_maxchiral},\eqref{trans_main}-\eqref{chiral anapoles_main} are important results of the present Letter. 
In this connection, notice that the conclusions  of Ref.~\cite{fernandez2016objects} particularized, on the one hand,  to magnetic scatterers, described by $\epsilon$ and $\mu$, and, on the other hand, to magneto-electric dipolar particles, are consistent with our results~\footnote{An important insight that we get within the current approach is that we have not imposed transparency to the opposite helicity of the incident field, $\sigma' = -\sigma$. In other words, we have not imposed $| \E^{\sigma -\sigma }_{\rm{sca}}( \r) | = 0$ to arrive to total transparency. In contrast, we have just required transparency in the same helicity channel of the incident field ($\sigma = \sigma'$), finding that the latter is  sufficient for total transparency to a given helicity of the incident field as extinction vanishes.}.

To get a visual insight into these effects, Fig.~\ref{F_1} depicts the scattering features of a  sphere of MECh under the illumination of an incident field with well-defined helicity.

At this stage, we establish the solutions to Eq.~\eqref{chiral anapoles_main}. That is, we will find the spectral trajectories at which spheres can yield MECh, in terms of a Riccati-Bessel function. Moreover,  we will also unravel the first Kerker condition of zero backscattering, $\tilde{a}_{\ell,\sigma} = \tilde{b}_{\ell,\sigma}$, in terms of a Riccati-Bessel function.


Spheres of MECh satisfy Eq.~\eqref{chiral anapoles_main}. However, it is challenging to find analytical solutions to Eq.~\eqref{chiral anapoles_main} when several multipolar orders contribute to the optical response of a chiral sphere. Then,  let us start this section by imposing Eq.~\eqref{chiral anapoles_main} for a fixed multipolar order $\ell$.  Now, by inspecting Eqs.~\eqref{a_l}-\eqref{c_l} and Eqs.~\eqref{primero}-\eqref{ultimo}, we notice that 
\begin{equation}  \label{OMEC_1}
a_\ell = b_\ell  = i \sigma c_\ell  \Longleftrightarrow
\left\{
    \begin{array}{lr}
        A_\ell (R) = \sigma B_\ell (R).\\
        A_\ell (L) = -\sigma B_\ell (L). 
    \end{array}
\right\}
\end{equation}
Equation~\eqref{OMEC_1} represents an important simplification to find the analytical condition ruling spheres of  MECh. Note that $W_\ell (J)$ and $V_\ell (J)$ for $J =L, R$ do not play any role in finding a sphere of MECh.
Moreover, it is vital to notice that $A_\ell (R) = \sigma B_\ell (R)$ and $A_\ell (L) = -\sigma B_\ell (L)$ must be satisfied simultaneously and independently. In other words, such spheres depend on the  helicity of the incident field. With this information in mind, we can now simplify Eq.~\eqref{OMEC_1} to
\begin{equation}  \label{OMEC_2}
\boxed{
a_\ell = b_\ell  = i \sigma c_\ell \Longleftrightarrow
    \begin{array}{lr}
        \psi_\ell (m_R x)\psi'_\ell ( x) =- \sigma \psi_\ell ( x) \psi'_\ell (m_R x). \\
        \psi_\ell (m_L x)\psi'_\ell ( x) =  \sigma \psi_\ell ( x) \psi'_\ell (m_L x).
    \end{array}
}
\end{equation}
Equation~\eqref{OMEC_2} is one of the primary key results of this Letter, as  it can be used as a general road map for finding the spectral trajectories at which chiral spheres can behave as objects of MECh. 
It is essential to emphasize that Eq.~\eqref{OMEC_2} is exact and, thus, remains valid regardless of the multipolar order $\ell$, refractive index contrast $m$, and optical size $x$ of the chiral sphere. Moreover, it is also crucial to note that we have not imposed any restriction on  the intrinsic chirality of the sphere, $\chi$. Hence, Eq.~\eqref{OMEC_2} could, in principle, be satisfied for an infinity of values of the set $\{\sigma, \ell, m, x, \chi \}$, underscoring the generality of our analytical derivation. 

At this point, we should also  highlight another important aspect included in Eq.~\eqref{OMEC_2}: The Kerker condition, K1,   given \bb by $\tilde{a}_{\ell,\sigma} = \tilde{b}_{\ell,\sigma}\; \forall \ell$~\cite{kerker1983electromagnetic, nieto2011angle, person2013demonstration, carretero2019kerker}. Notice that at K1,  the electromagnetic helicity is  preserved  since  $| \E^{\sigma -\sigma }_{\rm{sca}}( \r) | = 0$ (see Appendix~\ref{th}). We also note from Eq.~\eqref{OMEC_2} that  $\tilde{a}_{\ell,\sigma} = \tilde{b}_{\ell,\sigma} \; \forall \ell$ is a necessary  yet insufficient condition  for chiral spheres to be objects of  MECh.
Now, taking into account the effective relations of Mie coefficients, which are given by Eq.~\eqref{effective}, we infer that $\tilde{a}_{\ell,\sigma} = \tilde{b}_{\ell,\sigma} \;  \Longleftrightarrow  {a}_\ell = {b}_\ell \;$. That is, the chiral coefficient $c_\ell$ and the incident helicity of the field $\sigma$ do not play any role in the emergence of K1 in the scattering by chiral spheres.   Taking this fact into account, we can now manipulate  Eq.~\eqref{OMEC_2} to obtain after some algebra
\begin{equation}  
\boxed{
a_\ell = b_\ell  \Longleftrightarrow \psi_\ell (m_R x)\psi'_\ell ( m_L x) + \psi_\ell ( m_L x) \psi'_\ell (m_R x) = 0. \label{conservation}}
\end{equation}
Equation~\eqref{conservation} gives the spectral trajectories at which K1 is satisfied. Again, we remark  that no restrictions have been imposed on  Bohren´s exact solution ~\cite{bohren1974light}. Hence, Eq.~\eqref{conservation} is exact and remains valid for finding the first Kerker condition of zero backscattering, given by $a_\ell = b_\ell$, as a function of   $\{\ell, m, x, \chi \}$.

Hitherto, we have discussed spheres of MECh from the point of view of energy conservation. Later on, we found the exact analytical expression of  such spheres  in terms of a Riccati-Bessel function, inferred from Eq.~\eqref{OMEC_2}. Moreover,  from Eq.~\eqref{conservation}, we derived the first Kerker condition of zero backscattering in terms of the same Riccati-Bessel function. However, up to now, we have not explicitly shown the solutions to Eqs.~\eqref{OMEC_2}-\eqref{conservation}.

\begin{figure}[t!]
  \includegraphics[width=\columnwidth]{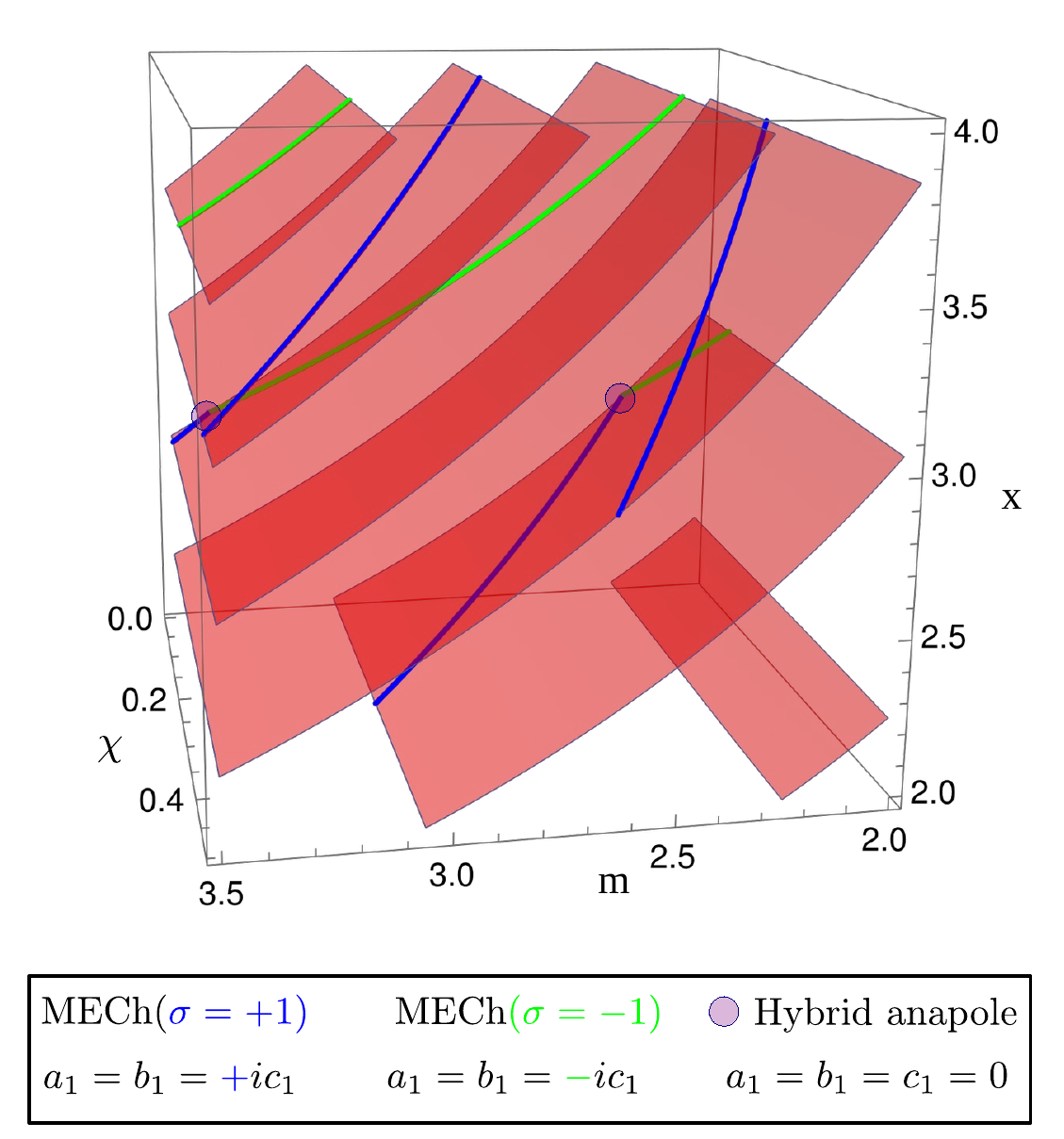}
  \caption{\color{black}{Chiral spheres of high refractive index: First Kerker condition,  Eq.\rr~\eqref{conservation} \bb,  given by $a_1 = b_1$ (red surfaces), and  spheres of MECh, given by $a_1 = b_1 = \im \sigma c_1$, (i.e., only   $\ell = 1$ contributes), as a function of the optical size of the chiral sphere $x$, refractive index contrast $m$, and intrinsic  chirality $\chi$.  The helicities of the incident field are given by $\sigma = 1$ (blue curves) and $\sigma = -1$ (green curves), respectively.   Purple circles denote hybrid anapoles given by $a_1 = b_1 = c_1 = 0$. }}
  \label{F_2}
\end{figure}

Figure~\ref{F_2}  shows  surfaces at which K1, $a_1 = b_1$, holds versus the optical size $x$, refractive index contrast $m$, and intrinsic chirality $\chi$ of a HRI  chiral sphere. Note that there are several sheets   that emerge from  Eq.~\eqref{conservation} for $l=1$ as  the  Riccati-Bessel functions $\psi_\ell  $ and $\psi_\ell ^{'} $   oscillate with   $x$ and $m$, while their  variation with   $\chi$ is very slow within its range of values.  

In addition, we show the spectral trajectories of MECh for incident helicity $\sigma = +1$ (blue curves) and $\sigma = -1$ (green curves). In this regard, it is essential to notice that two experiments are embedded in Fig.~\ref{F_2}: one with $\sigma = 1$ (green curves vanish) and the other with $\sigma = -1$ (blue curves vanish).

\begin{figure}[t!]
  \includegraphics[width=\columnwidth]{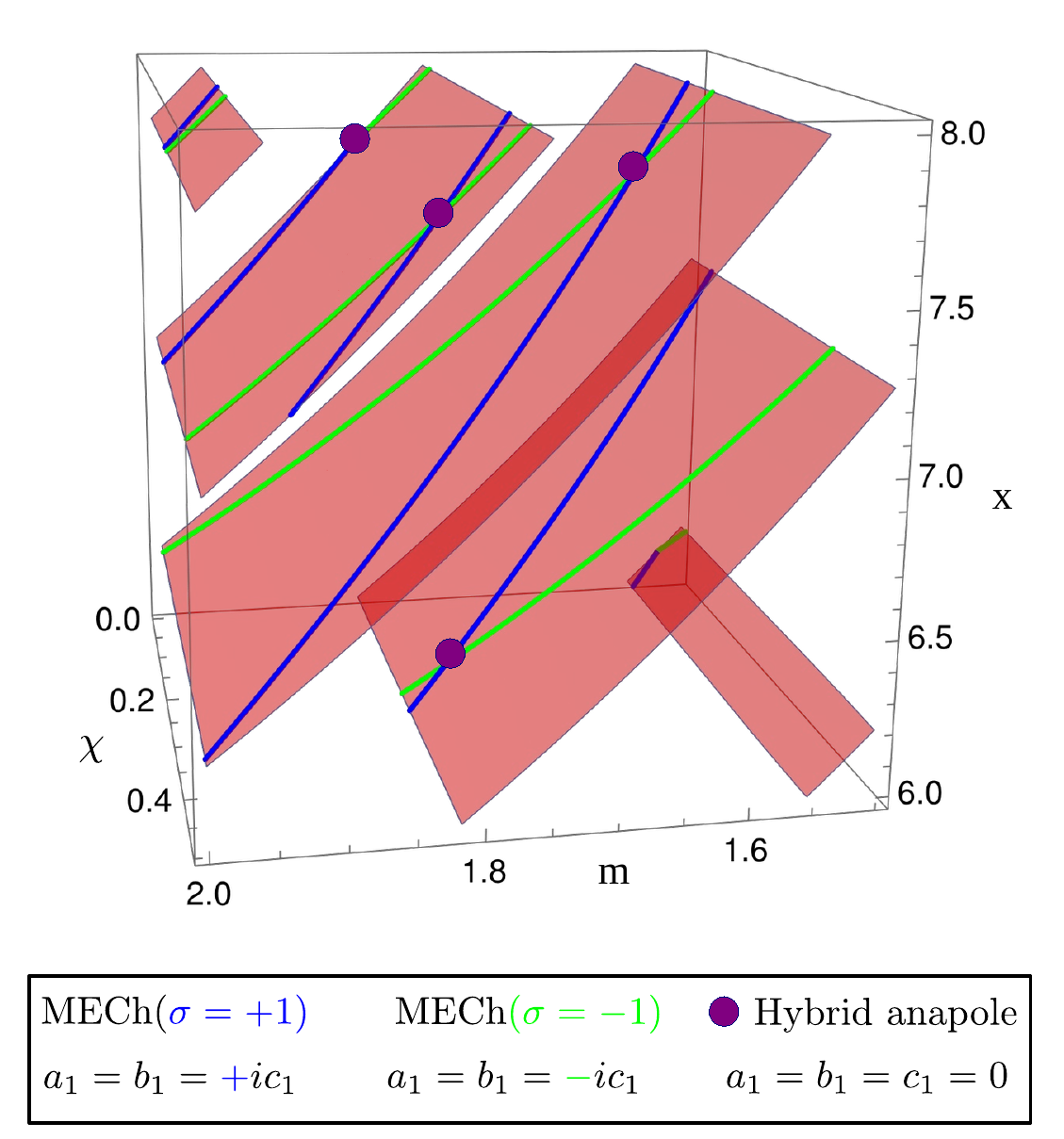}
  \caption{\color{black}{Chiral spheres of low refractive index: First Kerker condition, Eq.\rr~\eqref{conservation}\bb,  given by $a_1 = b_1$ (red surfaces), and  spheres of MECh, given by $a_1 = b_1 = \im \sigma c_1$, (i.e., only   $\ell = 1$ contributes) as a function of the optical size of the chiral sphere $x$, refractive index contrast $m$, and intrinsic  chirality $\chi$. The helicities of the incident field are given by $\sigma = 1$ (blue curves) and $\sigma = -1$ (green curves), respectively. Purple circles denote hybrid anapoles given by $a_1 = b_1 = c_1 = 0$. }}
  \label{F_3}
\end{figure}

We should acknowledge that,  as mentioned before and to the best of our knowledge, chiral HRI spheres have not yet been synthesized. To encourage experiments at the current state-of-the-art, Fig.~\ref{F_3} shows that spheres may exhibit  MECh  at lower refractive index contrasts ($1.5 <m <2$) for larger optical sizes ($6< x <8$). This range of parameters is of utmost current interest because previous experimental studies on chiral microspheres with low refractive indices have been reported in the literature~\cite{cipparrone2011chiral, donato2014polarization, shi2020chirality, venkatakrishnarao2017chiral}. 

Last but not least, let us briefly discuss the common features of  Figs.~\ref{F_2}-\ref{F_3}. First, we can note that a sphere of MECh surfs the first Kerker condition of zero backscattering for both helicities of the incident field. That is, the blue and green lines sail the red surfaces in Figs.~\ref{F_2}-\ref{F_3}.  In this regard, we note that these trajectories of MECh predominantly vanish upon flipping the helicity of the incident field from $\sigma \Longleftrightarrow -\sigma$. In other words, the blue and green lines do not generally intersect. However, this is not a general feature (see the purple circles).

In the following, we provide a physical explanation of these points. First,  we can infer that the crossing of spheres of MECh for $\sigma = +1$ and $\sigma = -1$ arises whenever $a_\ell = b_\ell = c_\ell = 0$. In fact, the purple circle represents a chiral sphere that behaves as a non-radiating source for both helicities of the incident field. In other words, a chiral sphere satisfying
$a_\ell = b_\ell = c_\ell = 0$ meets the hybrid anapole condition~\cite{luk2017hybrid, sanz2021multiple, coe2022unraveling}. 
In short, an hybrid anapole emerges when the electric and magnetic scattering coefficients simultaneously vanish regardless of the incident helicity. In Refs~\cite{luk2017hybrid,  sanz2021multiple, coe2022unraveling}, all illuminated samples are achiral ($\chi = 0$) and satisfy the hybrid anapole condition $a_\ell = b_\ell = 0$. In our case, we have a very different scenario:
from the purple circles in Figs.~\ref{F_2}-\ref{F_3}, we can infer that $\chi \neq 0$. That is, even if the sphere is chiral, it yields a scattered field that is identical to zero for both helicities. This fact arises since $c_\ell = 0$. In short, we can conclude that $\chi = 0 \rightarrow c_\ell = 0$ while $c_\ell = 0$ does not generally imply $\chi = 0$.

In conclusion, we have demonstrated  that chiral spheres may be objects yielding  maximum electromagnetic chirality.  That is, we have analytically found spectral trajectories in which chiral spheres are optically transparent to a given helicity of the incident field,  while for its opposite helicity, they behave as dual objects as a result of reciprocity.

In addition, we have unveiled the first Kerker condition of zero backscattering in terms of a Riccati-Bessel function. We recall that this condition preserves the incident helicity upon interaction, yielding zero backscattered light for cylindrically symmetry targets such as chiral spheres~\cite{zambrana2013duality}. 
In this connection, note that the so-called generalized Kerker effects~\cite{liu2018generalized,alaee2015generalized, bag2018transverse, shamkhi2019transverse}, which are other important cases of directional scattering for higher multipolar orders, do not generally correspond to dual particles, and thus, have not been addressed here.

Our results are exact and remain valid regardless of multipole order, refractive index, optical size, and intrinsic chirality of the  sphere.  Therefore, our  predictions can be used in experiments aimed to improve  dichroism characterization  and  the optical manipulation and sorting  of present-day low-refractive index chiral spheres. 

Additionally, our findings are of use to develop  building blocks for novel  maximum electromagnetic chirality nanoantennas, metasurfaces,  or metamaterials, encouraging experiments, and inspiring the fabrication of high refractive index chiral magnetoelectric spheres.

JO-T wishes to acknowledge his former thesis supervisor, Prof. Juan José Sáenz (Mole), for sharing his passion for science and helping him to become an independent physicist. He also thanks Jon Lasa-Alonso for initial conversations and Adrian Juan-Delgado for valuable insights. 
He also would like to thank Dr.~Cristina Sanz-Fernández for her helpful quantum comments.
 J.O-T. acknowledges support from the Juan de la Cierva fellowship No. FJC2021-047090-I of  MCIN/AEI/10.13039/501100011033 and NextGenerationEU/PRTR.  Research financial aid from this ministry is  also thanked by MN-V.  

\clearpage

\bibliography{Bib_tesis}
\clearpage

\newpage

\appendix

\section{Theoretical framework} \label{th}
Next, we lay out the framework to find that chiral spheres can be objects of MECh.  To that end,  let us  consider an incident field with well-defined helicity $\sigma = \pm 1$~\cite{olmos2019sectoral}, 
\be
\E^{\sigma}_{\text{inc}} ( \r) &=&  \sum_{\ell =1}^{\infty} \sum_{m=-\ell }^{+\ell }  C_{\ell m}^{ \sigma} \boldsymbol{\Psi}_{\ell m}^{\sigma} (\r).
\label{multipolar} 
\ee
Here $C_{\ell m}^{ \sigma}$ denotes the incident coefficients characterizing the nature of the wave, $k$ is the radiation wavevector, and 
\begin{subequations}
\be
\boldsymbol{\Psi}_{\ell m}^{\sigma} &=& \frac{1}{\sqrt{2}} \left[ {\boldsymbol{N}}_{\ell m} +  \sigma {\boldsymbol{M}}_{\ell m}  \right], \label{V1} \\
{\boldsymbol{M}}_{\ell m} &\equiv & j_\ell (kr)\boldsymbol{X}_{\ell m},  \quad
{\boldsymbol{N}}_{\ell m} \equiv \frac{1}{k} \boldsymbol{\nabla} \times {\boldsymbol{M}}_{\ell m}, \\
\boldsymbol{X}_{\ell m} &\equiv& \frac{1}{\sqrt{\ell (\ell +1)}} {\bf{L}} Y_{\ell m} (\theta,\varphi). \label{V3}
\ee
\end{subequations}
Here $\boldsymbol{M}_{\ell m}$ and $\boldsymbol{N}_{\ell m}$  are Hansen's multipoles~\cite{jackson1999electrodynamics}, $ j_\ell (kr)$ are the spherical Bessel functions,
$Y_{\ell m} (\theta,\varphi)$ are spherical harmonics, $\theta$ and $\varphi$ being  the polar and azimuthal angles,  and $ {\bf{L}} =  \left\{ -\im \r \times \boldsymbol{\nabla}\right\} $ represents the total angular momentum operator.

Let us expand the electric field scattered by a chiral sphere.
The Generalized Lorentz-Mie theory (GLMT) gives the exact solution to Maxwell's
equations for spherical particles (both chiral and achiral) in a homogeneous medium under general illumination conditions~\cite{gouesbet2011generalized}. Hence, we can employ the GLMT for the particular case in which an incident field with  well-defined helicity illuminates a chiral sphere. In the same basis of multipoles of well-defined helicity, the electric field scattered by a chiral sphere reads 
\be\label{sca_1}
\E^{\sigma}_{\rm{sca}}(\r) &=& \sum_{\sigma'=\pm1} \E^{\sigma \sigma'}_{\rm{sca}}(\r), \\ 
\E^{\sigma \sigma '}_{\rm{sca}}( \r) &=&   \sum_{\ell =0}^\infty \sum_{m=-\ell }^{+\ell } D^{\sigma \sigma' }_{\ell m}\boldsymbol{\Phi}_{\ell m}^{\sigma'}(\r). \label{sca_2}
\ee
Here $D^{\sigma \sigma'}_{\ell m} =C^{\sigma}_{\ell m} \left( \tilde{a}_{\ell,\sigma}  + \sigma \sigma' \tilde{b}_{\ell,\sigma} \right)/{2}$, $\sigma' = \pm 1$, denotes the decomposition of the scattered field   into  left - and right-handed CPL waves, and $ \boldsymbol{\Phi}_{\ell m}^{\sigma}$ is defined like   in Eq.~\eqref{V1} with the substitution of $h_\ell $ in place of $j_\ell $, $j_\ell $ and $h_\ell $ being the spherical Hankel functions of first and second kind, respectively~\cite{jackson1999electrodynamics}. Moreover,  $\tilde{a}_{\ell,\sigma}  = a_\ell  + i \sigma d_\ell$ and $\tilde{b}_{\ell,\sigma}  = b_\ell  - i \sigma c_\ell $ denote the effective Mie coefficients~\cite{ali2020probing}, $a_\ell $, $b_\ell $, $d_\ell $, and $c_\ell $ being the
Mie coefficients of the scattered field by a chiral sphere~\cite{bohren2008absorption}. 

Chiral spheres are, by their very nature, reciprocal objects. Hence, chiral spheres fulfill the Onsager-like relations, namely, $d_\ell  = -c_\ell $~\cite{onsager1936electric}. Following this fact,  we can simplify the previous effective relations  to~\cite{ali2020probing}
\begin{align} \label{effective}
\tilde{a}_{\ell,\sigma}  = a_\ell  - i \sigma c_\ell,  && \tilde{b}_{\ell,\sigma}  = b_\ell  - i \sigma c_\ell .
\end{align}

We shall see in the next section that the relationships given by Eq.~\eqref{effective} have significant consequences concerning  MECh spheres.

\section{Chiral Mie coefficients} \label{Bohren}

In 1974 Craig Bohren analytically solved the constitutive relations of chiral and reciprocal media for the case of a  sphere~\cite{bohren1974light}. The exact results that he derived  are fully captured by  the scattering coefficients $a_\ell $, $b_\ell $, and $c_\ell $, which according to Ref.~\cite{bohren2008absorption} read: 
\begin{subequations}
\be \label{a_l}
a_\ell  &=& \Delta^{-1}_\ell  \left[  V_\ell (R)A_\ell (L) + V_\ell (L)A_\ell (R) \right], \\ \label{b_l}
b_\ell  &=& \Delta^{-1}_\ell  \left[ W_\ell (L)B_\ell (R) + W_\ell (R)B_\ell (L) \right], \\ \label{c_l}
c_\ell  &=& i \Delta^{-1}_\ell  \left[  W_\ell (R) A_\ell (L) - W_\ell (L)A_\ell (R) \right],
\ee
\end{subequations}
where $\Delta_\ell  =  W_\ell (L) V_\ell (R) + V_\ell (L)W_\ell (R)$,
with 
\begin{subequations}
\be \label{primero}
W_\ell (J) &=& m \psi_\ell (m_J x) \xi^{'}_\ell (x)- \xi_\ell (x) \psi^{'}_\ell (m_J x), \\
V_\ell (J) &=& \psi_\ell (m_J x) \xi^{'}_\ell (x)- m\xi_\ell (x) \psi^{'}_\ell (m_J x), \\
A_\ell (J) &=& m \psi_\ell (m_J x) \psi^{'}_\ell (x)- \psi_\ell (x) \psi^{'}_\ell (m_J x), \\
B_\ell (J) &=& \psi_\ell (m_J x) \psi^{'}_\ell (x)- m\psi_\ell (x) \psi^{'}_\ell (m_J x). \label{ultimo}
\ee
\end{subequations}
Here $J = L, R$, with $m_L  = m - \chi$ and  $m_R = m + \chi$, $m$ being the refractive index contrast  particle/surrounding medium,  $\chi$ is a  small parameter modeling the  intrinsic  chirality of the sphere, $x = ka$ is the  size parameter of the particle,  $a$ being its radius. $\psi_\ell  (z) = z j_\ell (z)$ and $\xi_\ell  (z) = z h_\ell (z)$ are Riccati-Bessel functions. 
\clearpage
\end{document}